\documentclass[12pt]{article}
\usepackage{jhep-mod}
\usepackage{bm}
\usepackage{amssymb, amsmath, amsthm, physics}
\usepackage{mathrsfs}
\usepackage{hyperref}
\usepackage{multicol}
\usepackage{float, array, tabu}
\usepackage[tableposition=top]{caption}
\usepackage{tocloft}
\usepackage[utf8]{inputenc}
\usepackage{enumerate}
\usepackage{bigints}
\usepackage{appendix}
\usepackage{graphicx}
\usepackage{float}
\usepackage{tikz}
\usepackage{setspace}
\usepackage{cancel}
\definecolor{purple}{rgb}{1,0,1}
\definecolor{lime}{HTML}{A6CE39} 

\newcommand{\blue}[1]{{\color{blue} #1}}

\definecolor{lime}{HTML}{A6CE39}
\newcommand{\orcidicon}{%
	\begin{tikzpicture}
	\draw[lime, fill=lime] (0,0) 
		circle [radius=0.16] 
		node[white] {{\fontfamily{qag}\selectfont \tiny ID}};
	\draw[white, fill=white] (-0.0625,0.095) 
		circle [radius=0.007];
	\end{tikzpicture}
	\hspace{-5mm}
}

\newcommand\orcidMatt{{\href{https://orcid.org/0000-0003-1088-6485}{\orcidicon}}}
\begin{document}
\def\O{{\mathcal{O}}}
\title{\vspace{-75pt}
\huge{
Counterexamples to the maximum force conjecture 
\\
}
}
\author{
\Large
Aden Jowsey {\sf{and}} Matt Visser\orcidMatt}
\affiliation{School of Mathematics and Statistics, Victoria University of Wellington, \\
\null\qquad PO Box 600, Wellington 6140, New Zealand.}
\emailAdd{aden.jowsey@sms.vuw.ac.nz}
\emailAdd{matt.visser@sms.vuw.ac.nz}
\parindent0pt
\parskip7pt

\abstract{

\noindent
Dimensional analysis shows that the speed of light and Newton's constant of gravitation can be combined to define a quantity $F_* = {c^4\over G_N}$ with the dimensions of force (equivalently, tension). 
Then in \emph{any} physical situation we \emph{must} have $F_\mathrm{physical} = f \; F_*$, where the quantity  $f$ is some dimensionless function of dimensionless parameters. 
In many physical situations explicit calculation yields $f= \O(1)$, and quite often $f \leq {1\over4}$. 
This has lead multiple authors to suggest a (weak or strong) maximum force/maximum tension conjecture. 
Working within the framework of standard general relativity, we will instead focus on \emph{counter-examples} to this conjecture, paying particular attention to the extent to which the counter-examples are physically reasonable.  
The various counter-examples we shall explore strongly suggest that one should not put too much credence into any universal maximum force/maximum tension conjecture. 
Specifically, fluid spheres on the verge of gravitational collapse will generically violate the weak (and strong) maximum force conjectures.
If one wishes to retain any general notion of ``maximum force'' then one will have to very carefully specify precisely which forces are to be allowed within the domain of discourse. 

\bigskip
\noindent
{\sc Date:} Wednesday 3 February 2021; Tuesday 2 March 2021; \LaTeX-ed \today


}

{\enlargethispage{50pt}
\vspace{-30pt}
\setlength{\cftbeforesecskip}{3pt}
\maketitle}
\def\tr{{\mathrm{tr}}}
\def\diag{{\mathrm{diag}}}
\def\H{{\scriptscriptstyle{\mathrm{H}}}}
\def\O{{\mathcal{O}}}
\def\lint{\hbox{\Large $\displaystyle\int$}} 
\def\hint{\hbox{\Huge $\displaystyle\int$}}  
\clearpage
\section{Introduction}\label{S:intro}

The maximum force/maximum tension conjecture was independently mooted some 20 years ago by Gary Gibbons~\cite{Gibbons2002} and Christoph Schiller~\cite{Schiller1997}.
At its heart one starts by noting that in (3+1) dimensions the quantity 
\begin{eqnarray}
    \label{eq:F_*} F_* =\frac{c^4}{G_N} \approx 1.2\times10^{44} \hbox{ N}
\end{eqnarray}
has the dimensions of force (equivalently, tension). Here $c$ is the speed of light in vacuum, and $G_N$ is Newton's gravitational constant. Thereby \emph{any} physical force can \emph{always} be written in the form 
\begin{equation}
F_\mathrm{physical} = f\; F_*,
\end{equation}
where the quantity $f$ is some dimensionless function of dimensionless parameters. 
In very many situations~\cite{Gibbons2002,Schiller1997,Barrow2014,Barrow2020} explicit calculations yield $f \leq {1\over4}$, though sometimes numbers such as $f\leq {1\over2}$ also arise~\cite{Yen2018}. Specifically, Yen Chin Ong \cite{Yen2018} formulated strong and weak versions of the conjecture:
 \begin{enumerate}
 \itemsep-1pt
\item {Strong form:} \quad $f\leq {1\over4}$.         
 \item {Weak form:} \quad $f = \O(1)$.
 \end{enumerate}
 Note that $F_* = E_\mathrm{Planck}/L_\mathrm{Planck}$ can also be interpreted as the Planck force, though it is not intrinsically quantum as the various factors of $\hbar$ cancel, at least in (3+1) dimensions. Furthermore it is sometimes interesting~\cite{Schiller2005} to note that the Einstein equations 
\begin{equation}
G_{ab}= 8\pi \; {G_N\over c^4}\; T_{ab},
\end{equation}
can be written in terms of $F_*$ as 
 \begin{equation}
T_{ab} = { F_*\over8\pi} \; G_{ab}.
\end{equation}
When recast in this manner, maximum forces conjectures have tentatively been related to Jacobson's entropic derivation of the Einstein equations~\cite{Jacobson-entropy}.

Considerable work has also gone into attempts at pushing various modifications of the maximum force conjecture beyond the framework of standard general relativity~\cite{Dabrowski2015,Bolotin2015}. 
Overall, while there is little doubt that the quantity $F_*$ is physically important, we feel that the precise status of the maximum force conjecture is much less certain, and is less than universal.

We shall investigate these conjectures within the context of standard general relativity, focussing on illustrative \emph{counter-examples} based on simple physical systems, analyzing the internal forces, and checking the extent to which the counter-examples are physically reasonable. 
Specifically, we shall consider static spherically symmetric fluid spheres~\cite{Delgaty, Rahman:2001, Martin:2003-a, Martin:2003-b, Boonserm:2005, Boonserm:2006-a, Boonserm:2006-b, Boonserm:2007-1, Boonserm:2007-2}, and investigate both radial and equatorial forces. We shall also include an analysis of the speed of sound, and the relevant classical energy conditions, specifically the dominant energy condition (DEC),
see~\cite{ECs, book, twilight, Visser:1997, Visser:1996, Cattoen:2006, LNP-survey, Martin-Moruno:2013-a, Martin-Moruno:2013-b, Martin-Moruno:2015}.
We shall see that even the most elementary static spherically symmetric fluid sphere, Schwarzschild's constant density star, raises significant issues for the maximum force conjecture.  Other models, such as the Tolman~IV solution and its variants are even worse.
Generically, we shall see that any prefect fluid sphere on the verge of gravitational collapse will violate the 
weak (and strong) maximum force conjectures.
Consequently, if one wishes to retain any truly universal notion of ``maximum force'' then one will at the very least have to very carefully delineate precisely which forces are to be allowed within the domain of discourse.

\enlargethispage{40pt}
\vspace{-15pt}
\section{Spherical symmetry}\label{C:SS}
\vspace{-10pt}
Consider spherically symmetric spacetime, with metric given in Schwarzschild curvature coordinates:
\begin{equation}
ds^2 = g_{tt} \; dt^2 + g_{rr} \; dt^2 + r^2 (d\theta^2 +\sin^2\theta\; d\phi^2).
\end{equation}
We do not yet demand pressure isotropy, and for the time being allow radial and transverse pressures to differ, that is $p_r \neq p_t$. 

Pick a spherical surface at some specified value of the radial coordinate $r$. Define
\begin{equation}
    F_r(r) = \int p_r(r) \; dA = 4\pi \;p_r(r) \; r^2.
    \label{eq:Radial Force}
\end{equation}
This quantity simultaneously represents the compressive force exerted by outer layers of the system on the core, and the supporting force exerted by the core on the outer layers of the system. 

\enlargethispage{20pt}
Consider any equatorial slice through the system and define the equatorial force by
\begin{equation}
    F_{eq} = \int p_t(r) \; dA = 2\pi \int^{R_s}_{0} \sqrt{g_{rr}} \; p_t(r) \; r dr.
    \label{eq:Equatorial Force}
\end{equation}
This quantity simultaneously represents the force exerted by the lower hemisphere of the system on the upper hemisphere, and the force exerted by the upper hemisphere of the system on the lower hemisphere. 
Here $R_s$ is the location of the surface of the object (potentially taken as infinite). As we are investigating with spherically symmetric systems, the specific choice of hemisphere is irrelevant.

\section{Perfect fluid spheres}\label{C:PF}
\subsection{Generalities}\label{SS:generalities}

The perfect fluid condition excludes pressure anisotropy so that radial and transverse pressures are set equal: $p(r) = p_r(r) = p_t(r)$. Once this is done, the radial and equatorial forces simplify
\begin{equation}
    F_r(r) = \int p(r) \; dA = 4\pi \;p(r) \; r^2;
    \label{eq:Radial Force2}
\end{equation}
\begin{equation}
    F_{eq} = \int p(r) \; dA = 2\pi \int^{R_s}_{0} \sqrt{g_{rr}} \; p(r) \; r dr.
    \label{eq:Equatorial Force2}
\end{equation}
Additionally, we shall impose the conditions that pressure is positive and decreases as one moves outwards with zero pressure defining the surface of the object~\cite{Delgaty, Rahman:2001, Martin:2003-a, Martin:2003-b, Boonserm:2005, Boonserm:2006-a, Boonserm:2006-b, Boonserm:2007-1, Boonserm:2007-2}.\footnote{There is a minor technical change in the presence of a cosmological constant, the surface is then defined by $p(R_s) = p_\Lambda$.}
Similarly density is positive and does not increase as one moves outwards, though density need not be and typically is not zero at the surface~\cite{Delgaty, Rahman:2001, Martin:2003-a, Martin:2003-b, Boonserm:2005, Boonserm:2006-a, Boonserm:2006-b, Boonserm:2007-1, Boonserm:2007-2}.

We note that for the radial force we have by construction
\begin{equation}
F_r(0)=0; \qquad F_r(R_s) = 0; \qquad \hbox{and for} \quad r\in(0,R_s):\; F_r(r) > 0.
\end{equation}
In particular in terms of the central pressure $p_0$ we have the particularly simple bound
\begin{equation}
F_r(r) < 4\pi\; p_0 \; R_s^2.
\end{equation}
This suggests that in general an (extremely) weak version of the maximum force conjecture might hold for the radial force, at least within the framework outlined above, and as long as the central pressure is finite. Unfortunately without some general relationship between central pressure $p_0$ and radius $R_s$ this bound is less useful than one might hope. For the strong version of the maximum force conjecture no such simple argument holds for $F_r$, and one must perform a case-by-case analysis.
For the equatorial force $F_{eq}$ there is no similar argument of comparable generality, and one must again perform a case-by-case analysis.

Turning now to the classical energy conditions~\cite{ECs,book,twilight, Visser:1997, Visser:1996, Cattoen:2006, LNP-survey,Martin-Moruno:2013-a,Martin-Moruno:2013-b,Martin-Moruno:2015}, they add extra restrictions to ensure various physical properties remain well-behaved.  For our perfect fluid solutions, these act as statements relating the pressure $p$ and the density $\rho$ given by the stress-energy tensor $T_{\hat{\mu}\hat{\nu}}$.
Since, (in view of our fundamental assumptions that pressure and density are both positive), the null, weak, and strong energy conditions, (NEC, WEC, SEC) are always automatically satisfied, we will \emph{only} be interested in the dominant energy condition (DEC).
In the current context the dominant energy condition only adds the condition $|p| \leq \rho$. 
But since in the context of perfect fluid spheres, the pressure is always positive,  it is more convenient to simply write this as
\begin{equation}
    \frac{p}{\rho} \leq 1; \qquad \hbox{that is} \qquad p \leq \rho. 
    \label{eq:DEC}
\end{equation}
The best physical interpretation of the DEC is that it guarantees that any timelike observer with 4-velocity $V^a$ will observe a flux $F^a = T^{ab} \,V_b$ that is non-spacelike (either timelike or null)~\cite{LNP-survey}. However, it should be pointed out that the DEC, being the strongest of the classical energy conditions, is also the easiest to violate --- indeed there are several known situations in which the classical DEC is violated by quantum effects~\cite{book, twilight, Visser:1997, Visser:1996, Cattoen:2006, LNP-survey,Martin-Moruno:2013-a,Martin-Moruno:2013-b,Martin-Moruno:2015}.

The DEC is sometimes [somewhat misleadingly] interpreted in terms of the speed of sound not being superluminal: naively $v_s^2 = \partial p/\partial \rho \leq 1$; whence $p\leq \rho-\rho_\mathrm{surface} <\rho$. 
But the implication is only one-way, and in addition the argument depends on extra technical assumptions to the effect that the fluid sphere is well-mixed with a unique barotropic equation of state $p(\rho)$ holding throughout the interior. To clarify this point, suppose the equation of state is not barotropic, so that $p=p(\rho, z_i)$, with the $z_i$ being some collection of intensive variables,  (possibly chemical concentrations, entropy density, or temperature). Then we have
\begin{equation}
{dp \over dr } = {\partial p\over\partial\rho}\; {d\rho\over dr} + 
\sum_i {\partial p\over\partial z^i} \;{d z^i\over dr}
=
v_s^2(\rho,z^i)\; {d\rho\over dr} + 
\sum_i {\partial p\over\partial z^i} \;{d z^i\over dr}. 
\end{equation}
Then, (noting that $d\rho/dr$ is non-positive as one moves outwards), enforcing the speed of sound to not be superluminal implies
\begin{equation}
{dp \over dr } \geq {d\rho\over dr} + 
\sum_i {\partial p\over\partial z^i} \;{d z^i\over dr}. 
\end{equation}
Integrating this from the surface inwards we have
\begin{equation}
p(r) \leq \rho(r) - \rho(R_s)+ 
\sum_i  \int_r^{R_s} {\partial p(\rho,z^i)\over\partial z^i} \;{d z^i\over dr} \;dr.
\end{equation}
Consequently, unless one either makes an explicit barotropic assumption $\partial p/\partial z^i=0$, or otherwise at the very least has some very tight control over the partial derivatives $\partial p/\partial z^i$,
one simply cannot use an assumed non-superluminal speed of sound to deduce the DEC. 
Neither can the DEC be used to derive a non-superluminal speed of sound, at least not without many extra and powerful technical assumptions. We have been rather explicit with this discussion since we have seen considerable confusion on this point. 
Finally we note that there is some disagreement as to whether or not the DEC is truly fundamental~\cite{twilight, Visser:1997, Visser:1996, Cattoen:2006}.

\subsection{Schwarzschild's constant density star}
We shall now consider a classic example of perfect fluid star, Schwarzschild's constant density star~\cite{Sch:Int}, (often called the Schwarzschild interior solution), which was discovered very shortly after Schwarzschild's original vacuum solution~\cite{Sch:Ext}, (often called the Schwarzschild exterior solution).  

It is commonly argued that Schwarzschild's constant density star is ``unphysical'' on the grounds that it allegedly leads to an infinite speed of sound. But this is a naive result predicated on the physically unreasonable hypothesis that the star is well-mixed with a barotropic equation of state $p=p(\rho)$. To be very explicit about this, all realistic stars are physically stratified with non-barotropic equations of state $p=p(\rho, z_i)$, with the $z_i$ being some collection of intensive variables,  (possibly chemical concentrations, entropy density, or temperature). We have already seen that
\begin{equation}
{dp \over dr } = {\partial p\over\partial\rho}\; {d\rho\over dr} + 
\sum_i {\partial p\over\partial z^i} \;{d z^i\over dr}
=
v_s^2(\rho,z^i)\; {d\rho\over dr} + 
\sum_i {\partial p\over\partial z^i} \;{d z^i\over dr}. 
\end{equation}
Thence for a constant density star, $d\rho/dr=0$, we simply deduce
\begin{equation}
{dp \over dr } = 
\sum_i {\partial p\over\partial z^i} \;{d z^i\over dr}. 
\end{equation}
This tells us nothing about the speed of sound, one way or the other --- it does tell us that there is a fine-tuning between the pressure $p$ and the intensive variables $z^i$, but that is implied by  the definition of being a ``constant density star''. 
We have been rather explicit with this discussion since we have seen considerable confusion on this point. Schwarzschild's constant density star is not ``unphysical''; it may be ``fine-tuned'' but it
 is not \emph{a priori} ``unphysical''.

\clearpage
Specifically, the Schwarzschild interior solution describes the geometry inside a static spherically symmetric perfect fluid constant density star with radius $R_s$ and mass $M$ by the metric:
\begin{equation}
\label{Eq:IntSchwarzschild}
ds^2 = -\frac{1}{4}\left(3\sqrt{1- \frac{2 M}{ R_s}} - \sqrt{1 - \frac{2 M r^2}{ R_s^3}}\right)^2 dt^2 
+ \left(1 - \frac{2 M r^2}{ R_s^3}\right)^{-1} dr^2 + r^2d\Omega^2.
\end{equation}
Here we have adopted geometrodynamic units ($c\to1$, $G_N\to1$).  Calculating the non-zero orthonormal stress-energy components from the Einstein equations applied to this metric yields: 
\begin{eqnarray}
    T_{\hat{t}\hat{t}} &=& \rho =  \frac{3M}{4\pi R_s^3};\\
    T_{\hat{r}\hat{r}} &=& T_{\hat{\theta}\hat{\theta}} =   T_{\hat{\phi}\hat{\phi}}
    = p =    
    \rho \;\; \frac{\sqrt{1-\frac{2 M r^2}{ R_s^3}}-\sqrt{1-\frac{2 M}{ R_s}}}
    {3\sqrt{1-\frac{2 M}{ R_s}}-\sqrt{1-\frac{2 M r^2}{ R_s^3}}}.
 \end{eqnarray}
This gives us the relation between density and pressure, as well as demonstrating the perfect fluid condition ($p = p_r = p_t$), and also verifying that the density is (inside the star) a position independent constant. 
In these geometrodynamic units both density and pressure have units 1/(length)$^2$, while forces are dimensionless.
Note that the pressure does in fact go to zero at $r\to R_s$, so $R_s$ really is the surface of the ``star''. 
Rewriting the relation between pressure and density in terms of the simplified dimensionless quantities $\chi = \frac{2 M}{ R_s}$ and $y = \frac{r^2}{R_s^2}$ we see
\begin{eqnarray}
    \label{eq:InteriorPressure}
    p =  \rho \;\;\frac{\sqrt{1-\chi y}-\sqrt{1-\chi}}{3\sqrt{1-\chi}-\sqrt{1-\chi y}}.
\end{eqnarray}
Here $0 \leq y \leq 1$, and $0\leq \chi <\frac{8}{9}$. The first of these ranges is obvious from the definition of $y$, while the second comes from considering the central pressure at $y=0$:
\begin{equation}
p_0 = \rho \;\;\frac{1-\sqrt{1-\chi}}{3\sqrt{1-\chi}-1}.
\end{equation}
Demanding that the central pressure be finite requires $\chi <\frac{8}{9}$. 
(This is actually a rather more general result of general relativistic stellar dynamics, not restricted to constant density, see various discussions of the Buchdahl--Bondi bound \cite{Buchdahl-bound,Bondi-bound}.)

\subsubsection{Radial Force}

The radial force $F_r$ as defined by equation (\ref{eq:Radial Force2}) can be combined with the pressure-density relation given by equation (\ref{eq:InteriorPressure}), giving:
\begin{equation}
    \label{eq:ISC F_r}
    F_r = 4\pi p r^2 = 4\pi  \rho R_s^2\; y \; \frac{\sqrt{1-\chi y}-\sqrt{1-\chi}}{3\sqrt{1-\chi}-\sqrt{1-\chi y}} = \frac{3}{2} \chi y \; \frac{\sqrt{1-\chi y}-\sqrt{1-\chi}}{3\sqrt{1-\chi}-\sqrt{1-\chi y}}.
\end{equation}
As advertised in both abstract and introduction, this quantity is indeed a dimensionless function of dimensionless variables. Furthermore this quantity  is defined on the bounded range $0 \leq y \leq 1$, $0\leq \chi <\frac{8}{9}$. To find if $F_r$ itself is bounded we analyse the multi-variable derivative for critical points.

\enlargethispage{20pt}
For $\partial_\chi F_r$ we find:
\begin{equation}
\partial_\chi F_r = -{3 y\over2} \left(\frac
{ \{4-\chi(3+y) \} \sqrt{1-\chi}\sqrt{1-\chi y} -\{4-\chi(3+5y-4\chi y)\}  }
{\sqrt{1-\chi}\sqrt{1 - \chi y}\;(3\sqrt{1-\chi}-\sqrt{1 - \chi y})^2} \right).
\end{equation}
For $\partial_y F_r$ we find:
\begin{equation}
\partial_y F_r = -{3\chi\over2} \left(\frac
{\{4-\chi(3+y)\} \sqrt{1-\chi y}-\{4-5\chi y\} \sqrt{1-\chi} }
{\sqrt{1 - \chi y}\;(3\sqrt{1-\chi}-\sqrt{1 - \chi y})^2} \right).
\end{equation}
In particular we see that
\begin{equation}
\chi \partial_\chi F_r - y \partial_y F_r = 
{3\chi^2 y\over2} \left(\frac
{ \sqrt{1-\chi y}) }
{\sqrt{1-\chi}\;(3\sqrt{1-\chi}-\sqrt{1 - \chi y})^2} \right).
\end{equation}
To have a critical point, $ \partial_\chi F_r = \partial_y F_r = 0$, we certainly require $\chi y =0$. So either $\chi=0$ or $y=0$. 
But for $y=0$, and $\chi\in(0,{8\over9})$ we have
\begin{equation}
\partial_y F_r  
\quad\longrightarrow\quad 
 {3\chi(3\chi + 4 \sqrt{1-\chi}-4)\over 2 (3\sqrt{1-\chi}-1)^2} > 0. 
\end{equation}
In contrast, for $\chi=0$, and $y\in(0,1)$, we have $\partial_\chi F_r  \to 0$. 
So the only critical points lie on one of the boundary segments:
\begin{equation}
 \partial_\chi F_r = \partial_y F_r = 0 \quad\iff\quad \chi = 0.
\end{equation}
Therefore to find the maxima of $F_r(\chi,r)$ we must inspect all four of the boundary segments of the viable region. Along three of the boundary segments we can see that the three lines corresponding to $\chi =0$, $y = 0$, and $y = 1$ all give $F_r(\chi,r) = 0$, leaving only $\chi \rightarrow \frac{8}{9}$ to be investigated. 

We note
\begin{equation}
 \lim_{\chi \rightarrow \frac{8}{9}}  F_r(\chi, y) = {4y\over3}\; {(\sqrt{9-8y}-1)\over3-\sqrt{9-8y}}. 
\end{equation}
Inserting this into the partial derivative $\partial_y F_r$ reveals:
\begin{equation}
    \lim_{\chi \rightarrow \frac{8}{9}} \partial_y F_r = -\frac{4}{3} \left(1 + \frac{1}{\sqrt{9 - 8y}}\right).
\end{equation}
This is a strictly negative function in the range $0 \leq y \leq 1$. 

\begin{figure}[!!htbp]
\centering
\includegraphics{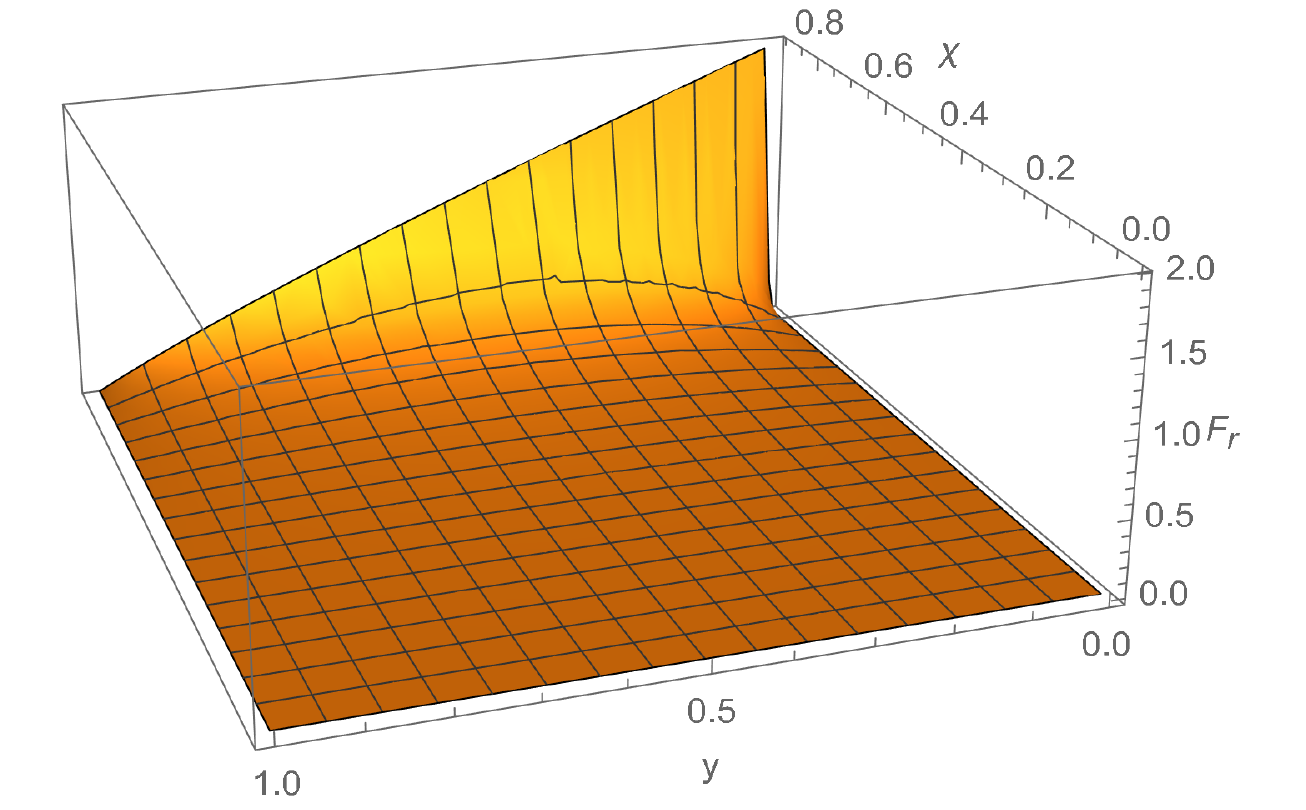}
\caption{Radial force $F_r(\chi,y)$ for the interior Schwarzschild solution. \hfill \break
Note $F_r(\chi,y)$ is bounded above by 2 in the region of interest $y\in[0,1]$, $\chi\in[0,8/9)$. }
\label{Fig:IntSch-Radial-Force}
\end{figure}

Thus the maximum of $F_r(\chi,y)$ can be found by taking the limit $\lim_{y\rightarrow 0}$ giving:
\begin{equation}
    (F_r)_\mathrm{max} = \lim_{y\rightarrow 0} \lim_{\chi \rightarrow \frac{8}{9}} F_r = 2.
\end{equation}
This is therefore bounded, with the radial force approaching its maximum at the centre of a fluid star which is on the verge of collapse. This force violates the strong maximum force conjecture, though it satisfies the weak maximum force conjecture. This limit can easily be seen graphically in Figure \ref{Fig:IntSch-Radial-Force}. 

\subsubsection{Equatorial force}

Using equation (\ref{eq:Equatorial Force2}) and the metric defined in equation (\ref{Eq:IntSchwarzschild}), with the relabelling of the previous subsection in terms of $\chi$ and $y$ gives:
\begin{eqnarray}
    F_{eq}(\chi) = \frac{3}{8}\;\chi\; \lint^1_{\!\!\!\!\!0} \frac{1}{\sqrt{1-\chi y}}\; \left(\frac{\sqrt{1-\chi y}-\sqrt{1-\chi}}{3\sqrt{1-\chi}-\sqrt{1-\chi y}}\right) dy. 
\end{eqnarray}
The integral evaluates to:
\begin{eqnarray}
    F_{eq}(\chi) &=& \frac{3}{4} \left.\left[\sqrt{1-\chi y} + 2\sqrt{1-\chi}\;\ln(3\sqrt{1-\chi} - \sqrt{1-\chi y}) \right]\right\vert_{y=0}^{y=1}.
\end{eqnarray}
Ultimately 
\begin{eqnarray}
\label{E:Feq-sch}
    F_{eq}(\chi)
        &=&\frac{3}{4}\left[\sqrt{1-\chi}\left\{1 + \ln(4-4\chi)-2\ln(3\sqrt{1-\chi}-1)\right\}-1 \right].
\end{eqnarray}
However, due to the presence of the $-\ln(3\sqrt{1-\chi}-1)$ term in this equation, it can be seen that as $\chi \rightarrow \frac{8}{9}$, $F_{eq}(\chi) \rightarrow +\infty$.
Indeed
\begin{equation}
F_{eq}(\chi) = {\ln({8\over9}-\chi)\over 2} + \O(1),
\end{equation}
implying that the equatorial force in this space-time will grow without bound as the star approaches the critical size, (just prior to gravitational collapse), in violation of both the strong and weak maximum force conjectures.

So while the interior Schwarzschild solution has provided a nice example of a bounded radial force, $F_r(y,\chi)$,  it also clearly provides an explicit  counter-example, where the equatorial force $F_{eq}(\chi)$ between two hemispheres of the fluid star grows without bound.

\subsubsection{DEC}
Imposing the DEC (equation \ref{eq:DEC}) within the fluid sphere we would require:
\begin{equation}
\frac{p}{\rho}=\frac{\sqrt{1-\chi y}-\sqrt{1-\chi}}{3\sqrt{1-\chi}-\sqrt{1-\chi y}}\leq1
\end{equation}
\enlargethispage{20pt}
That is
\begin{equation}
\sqrt{1-\chi y} \leq 4\sqrt{1-\chi},
\end{equation}
whence
\begin{equation}
1-\chi y \leq 16(1-\chi). 
\end{equation}

\begin{figure}[!htbp]
\centering
\begin{minipage}{.5\textwidth}
  \centering
  \includegraphics[width=1\linewidth]{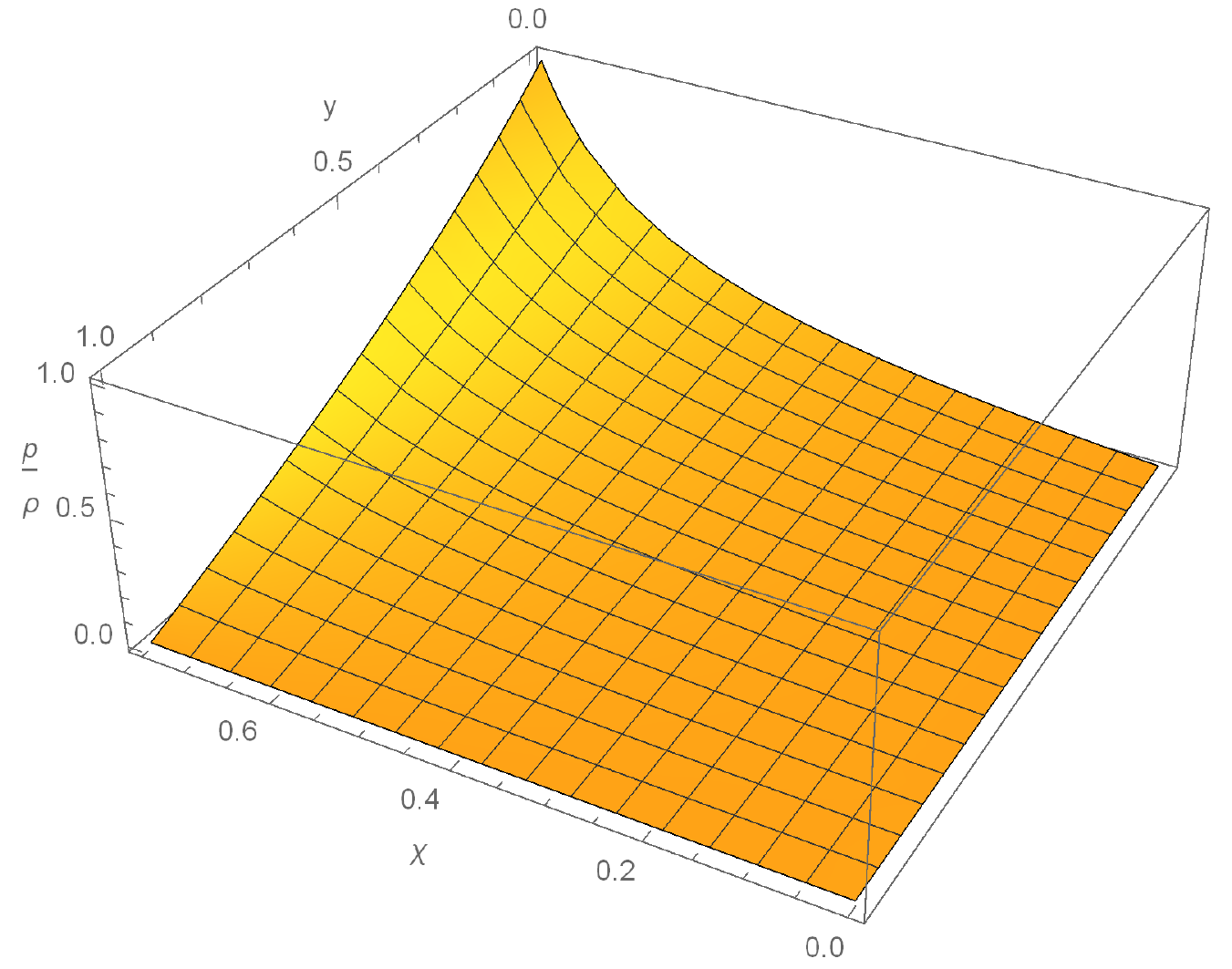}
  \caption{$\frac{p}{\rho}$ in first range}
  \label{F:IntSw-DEC1}
\end{minipage}%
\begin{minipage}{.5\textwidth}
  \centering
  \includegraphics[width=1\linewidth]{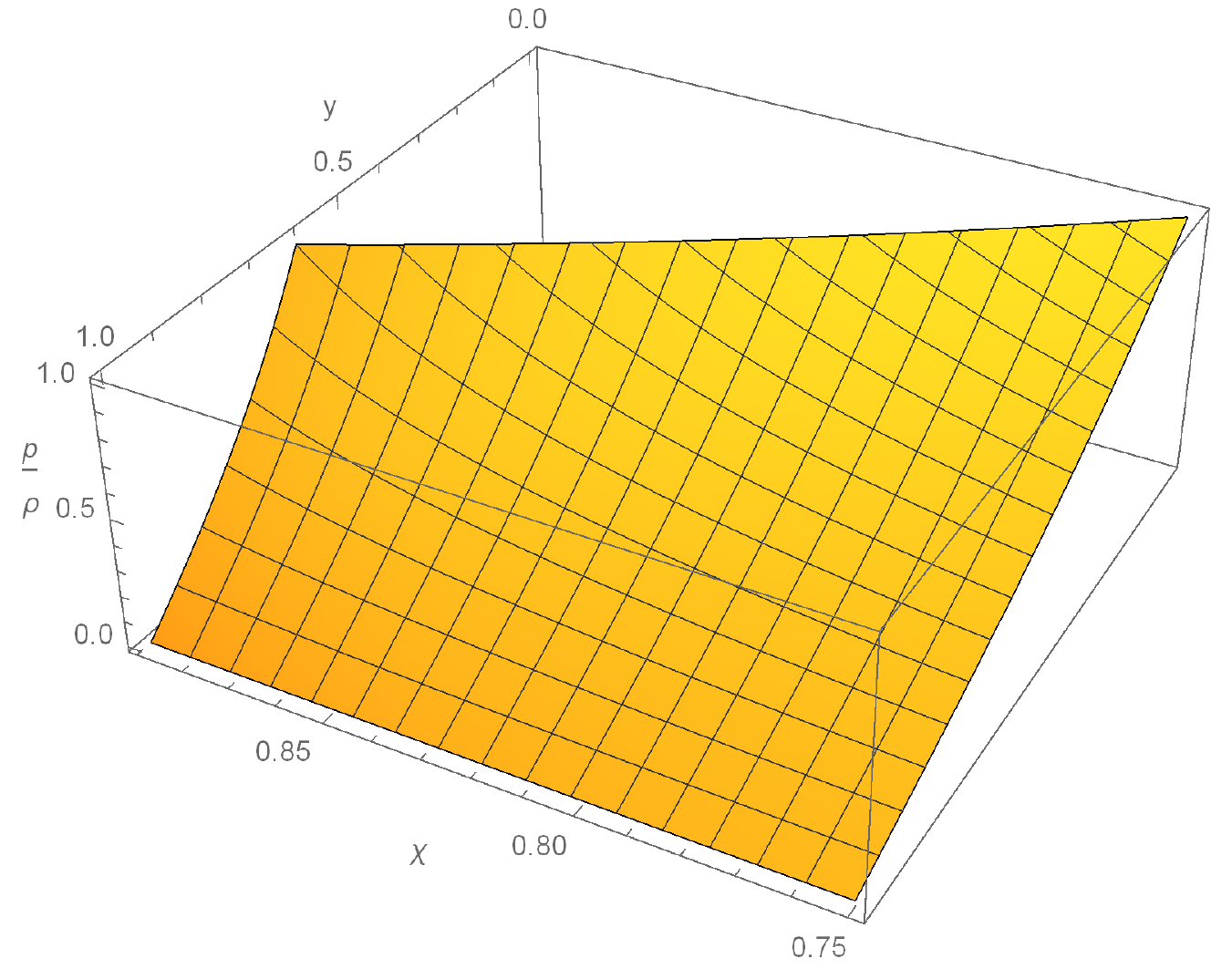}
  \caption{$\frac{p}{\rho}$ in second range}
  \label{F:IntSw-DEC2}
\end{minipage}
\end{figure}

Applying the boundary conditions of $0\leq\chi\leq\frac{8}{9}$, $0\leq y \leq 1$, we have a solution range:
\begin{equation}
    \left(0\leq \chi \leq \frac{3}{4},\quad 0\leq y \leq 1\right) 
    \quad\quad \bigcup \quad\quad 
    \left(\frac{3}{4}< \chi \leq \frac{8}{9},\quad 4 - \frac{3}{\chi}\leq y \leq 1\right).
\end{equation}
See figures \ref{F:IntSw-DEC1} and  \ref{F:IntSw-DEC2}.

\begin{figure}[!htbp]
\begin{center}
\includegraphics{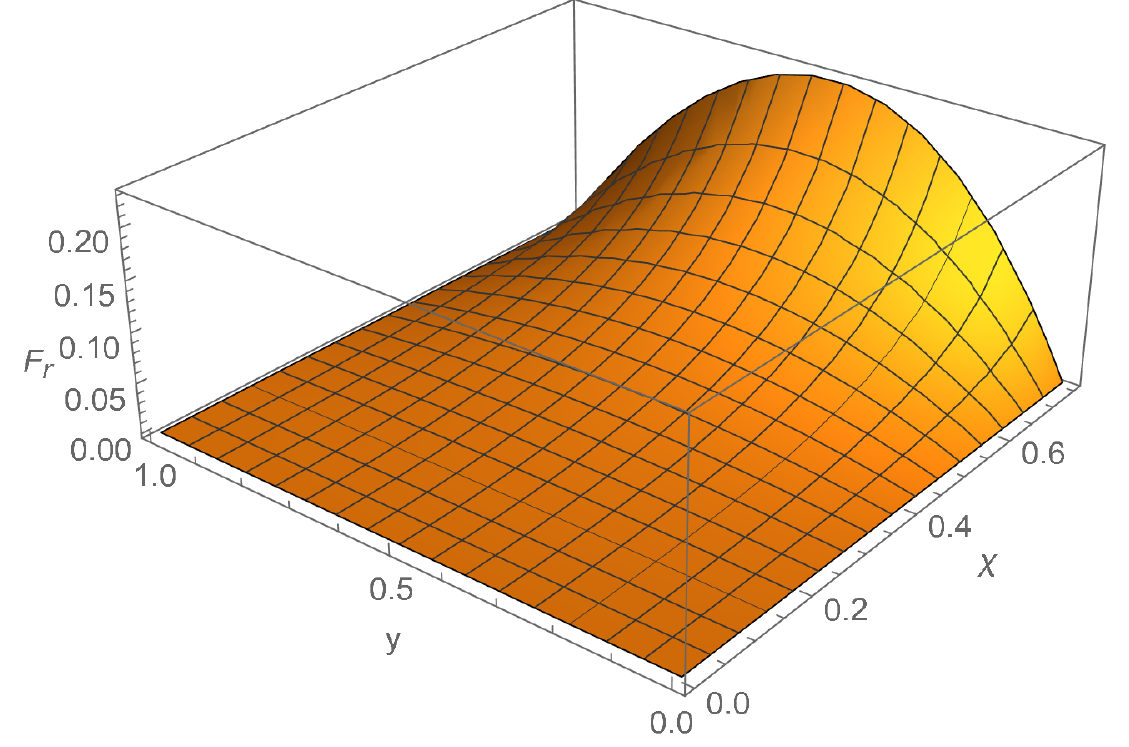}
\caption{Radial force $F_r(\chi,y)$ for the interior Schwarzschild solution in region 1 $ \left(0\leq \chi \leq \frac{3}{4},\quad 0\leq y \leq 1\right) $ where the DEC is satisfied.}
\label{F:region-1}
\end{center}
\end{figure}

Within the first region $0\leq \chi\leq \frac{3}{4},\quad 0\leq y\leq 1$, the radial force is maximised at:
\begin{equation}
\chi = \frac{3}{4},\quad y=\frac{1}{6}(5-\sqrt{5})\approx 0.46;
\quad\rightarrow \quad F_r=\frac{3}{16}(\sqrt{5}-1)\approx 0.23 <{1\over4}.
\end{equation}
Under these conditions the strong maximum force conjecture is satisfied.
This can be seen visually in figure~\ref{F:region-1}.

\begin{figure}[!htbp]
\begin{center}
\includegraphics{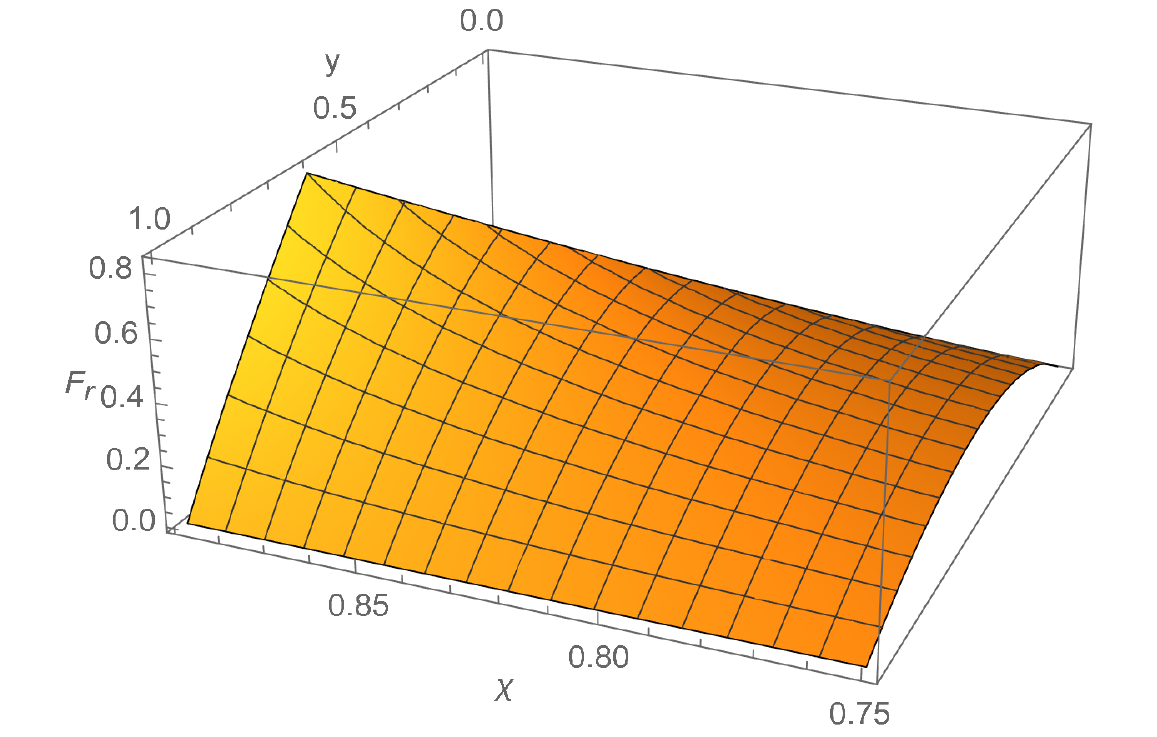}
\caption{Radial force $F_r(\chi,y)$ for the interior Schwarzschild solution in region 2 $\left(\frac{3}{4}< \chi \leq \frac{8}{9},\quad 4 - \frac{3}{\chi}\leq y \leq 1\right)$ where the DEC is satisfied.}
\label{F:region-2}
\end{center}
\end{figure}

Within the second region $\frac{3}{4}< \chi\leq \frac{8}{9},\quad 3-\frac{4}{\chi}\leq y\leq 1$, the radial force is maximised at:
\begin{equation}
\chi = \frac{8}{9},\quad y=\frac{5}{8};\quad\rightarrow \quad F_r=\frac{5}{6} > {1\over4}.
\end{equation}
Under these conditions the strong maximum force conjecture is violated, though the weak maximum force conjecture is satisfied.
This can be seen visually in figure~\ref{F:region-2}.

\enlargethispage{40pt}
Turning to the equatorial force, we see that the integrand used to define integral for $F_{eq}(\chi)$ satisfies the DEC only within the range $0\leq\chi\leq\frac{3}{4}$. Using the result for $F_{eq}(\chi)$ given above, equation (\ref{E:Feq-sch}), we have:
\begin{equation}
(F_{eq})_\mathrm{max,DEC}  = F_{eq}(\chi=3/4) = {3\over8} (2\ln 2 -1) \approx 0.1448603854 <{1\over4}.
\end{equation}
This now satisfies the strong maximum force conjecture.

\subsubsection{Summary}

Only if we enforce the DEC can we then make Schwarzschild's constant density star satisfy the weak and strong maximum force conjectures. Without adding the DEC Schwarzschild's constant density star will violate both the weak and strong maximum force conjectures. 
Since it is not entirely clear that the DEC represents fundamental physics~\cite{twilight, Visser:1997, Visser:1996, Cattoen:2006}, 
it is perhaps a little sobering to see that one of the very simplest idealized stellar models already raises issues for the maximum force conjecture.
We shall soon see that the situation is even worse for the Tolman~IV model (and its variants).

\subsection{Tolman IV solution}
The Tolman IV solution is another perfect fluid star space-time, however it does not have the convenient (albeit fine-tuned) property of constant density like the interior Schwarzschild solution. The metric can be written in the traditional form \cite{Delgaty}:
\begin{equation}
    ds^2 = -\left(1 + \frac{r^2}{A^2}\right)dt^2 +  \frac{1  + \frac{2r^2}{A^2}}{\left(1- \frac{r^2}{R^2}\right)\left(1+ \frac{r^2}{A^2}\right)}dr^2 + r^2d\Omega^2.
\end{equation}
Here $A$ and $R$ represent some arbitrary scale factors with units of length. This metric yields the orthonormal stress-energy tensor: 
\begin{eqnarray}
    T_{\hat{t}\hat{t}} &=& \rho =  \frac{1}{8\pi}\frac{6r^4 + (7A^2 + 2R^2) r^2 + 3A^2(A^2 + R^2)}{ R^2(A^2 + 2r^2)^2};\\
    T_{\hat{r}\hat{r}} &=& T_{\hat{\theta}\hat{\theta}} = T_{\hat{\phi}\hat{\phi}} = p =   \frac{1}{8\pi } \frac{R^2 - A^2 - 3r^2}{R^2(A^2 + 2r^2)}.
\end{eqnarray}
This demonstrates the non-constancy of the energy-density $\rho$ as well as the perfect fluid conditions. Physically, the surface of a fluid star is defined as the zero pressure point, which now is at:
\begin{equation}
    \label{eq:R_s}
    R_s = \sqrt{\frac{R^2 - A^2}{3}}.
\end{equation}
And likewise we can find the surface density ($\rho$ at $R_0$):
\begin{equation}
    \rho_s = \frac{3}{4\pi }\frac{2A^2 + R^2}{R^2(A^2 + 2R^2)}.
\end{equation}
The central pressure and central density are
\begin{equation}
p_0 = {1\over8\pi} {R^2-A^2\over R^2 A^2}; \qquad
\rho_0 ={1\over8\pi} {3(R^2+A^2)\over R^2 A^2}.
\end{equation}

Moving forwards, we will likewise calculate the radial and equatorial forces in this space-time.

\subsubsection{Radial force}
Using the previously defined radial force equation, (\ref{eq:Radial Force}), we can write the radial force for the Tolman IV space-time as:
\begin{equation}
    F_r(r,R,A) = 4\pi p_r r^2 
    = \frac{r^2}{2 R^2}  \left(\frac{R^2 - A^2 - 3r^2}{A^2 + 2r^2}\right).
\end{equation}
Defining $y= r^2/R_s^2$ and $a=A^2/R^2$ we have $y\in(0,1)$ and $a\in(0,1)$. The radial force then simplifies to
\begin{equation}
F_r(a,y) = {(1-a)^2 y (1-y)\over 6a + 4 (1-a) y}.
\end{equation}
Note this is, as expected, a dimensionless function of dimensionless variables.

\begin{figure}[!htbp]
\begin{center}
\includegraphics{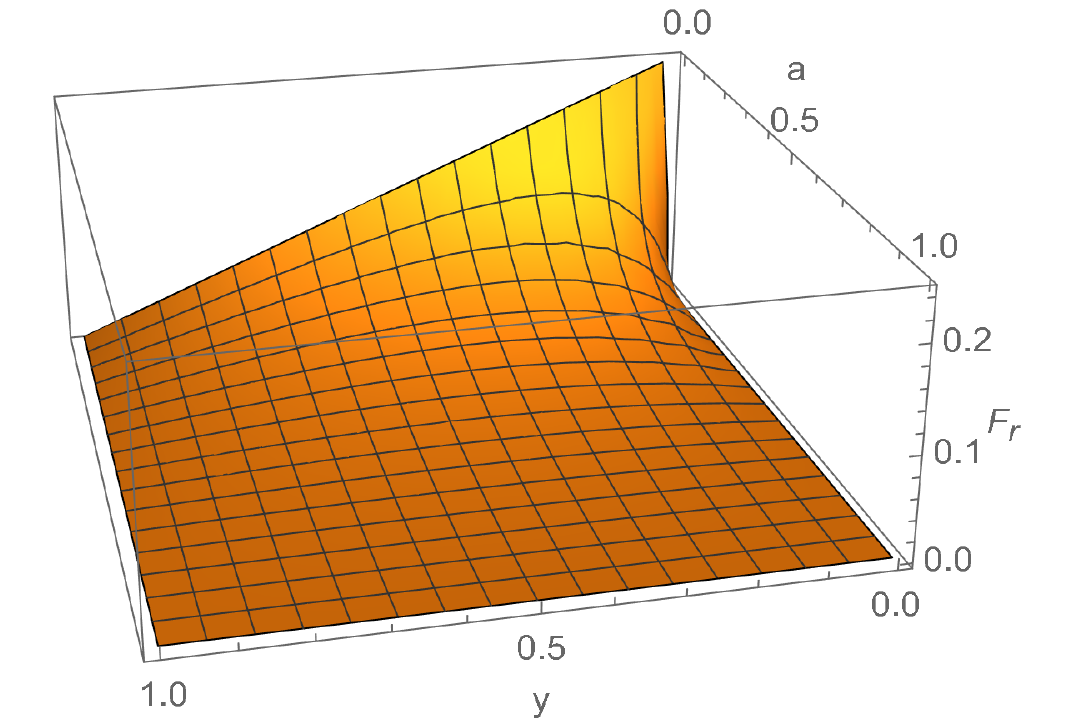}
\caption{$F_r(a,y)$ for the Tolman~IV solution.}
\label{F:Tolman-IV-Fr}
\end{center}
\end{figure}

The multivariable derivatives are:
\begin{equation}
\partial_a F_r(a,y)= {y(1-y)(1-a)(2ay-3a-2y-3)\over2(2ay-3a-2y)^2}.
\end{equation}
\begin{equation}
\partial_y F_r(a,y)=  {(1-a)^2(2ay^2-6ay-2y^2-3a)\over2(2ay-3a-2y)^2}.
\end{equation}
For both derivatives to vanish, (within the physical region), we require $a=1$. 
However $a=1$ actually minimises the function with $F_r(a,y)=0$. So we need to look at the boundaries of the physical region. Both $y=0$ and $y=1$ also minimise the function with $F_r(a,y)=0$. We thus consider $a=0$:
\begin{equation}
F_r(0,y)=\frac{1}{4}(1-y),
\end{equation}
where then it is clear that the function is maximised at $a=0$, $y\to0$, which corresponds 
$(F_r)_\mathrm{max}={1\over 4}$.
This can be seen visually in figure~\ref{F:Tolman-IV-Fr}.
Thus $F_r(a,y)$ for the Tolman~IV solution is compatible with the strong maximum force conjecture, but as for the Schwarzschild constant density star, we shall soon see that the equatorial force does not behave as nicely.

\subsubsection{Equatorial force}

Using equation (\ref{eq:Equatorial Force2}) for this space-time, and combining it with radial surface result of equation (\ref{eq:R_s}), we obtain:
\begin{equation}
    F_{eq} = 2\pi\int^{R_s}_0\sqrt{g_{rr}} \;p(r)\; r\, dr 
    = \frac{1}{4}\lint^{\sqrt{\frac{R^2 - A^2}{3}}}_{\!\!\!\!\!0}
    \sqrt{\frac{1  + \frac{2r^2}{A^2}}{\left(1- \frac{r^2}{R^2}\right)\left(1+ \frac{r^2}{A^2}\right)}} 
    \;  \frac{R^2 - A^2 - 3r^2}{R^2(A^2 + 2r^2)} \; r \,dr.
\end{equation}
As an integral this converges, however the resultant function is intractable. Instead, we will opt for a simpler approach by finding a simple bound. Since the radial coordinate is physically bound by $0\leq r \leq R_s = \sqrt{\frac{R^2 - A^2}{3}}<R$, we find that in that range:
\begin{eqnarray}
    g_{rr} = \frac{1}{1- \frac{r^2}{R^2}} \frac{1  + \frac{2r^2}{A^2}}{1+ \frac{r^2}{A^2}} \geq  \frac{1  + \frac{2r^2}{A^2}}{1+ \frac{r^2}{A^2}} \geq 1.
\end{eqnarray}
This is actually a much more general result; for any perfect fluid sphere we have
\begin{equation}
g_{rr}={1\over1-2m(r)/r},
\end{equation}
where $m(r)$ is the Misner--Sharp quasi-local mass. 

So as long as $m(r)$ is positive, which is guaranteed by positivity of the density $\rho(r)$, we have $g_{rr}>1$, and so in all generality we have 
\begin{equation}
F_{eq} >  2\pi\int^{R_s}_0 \;p(r)\; r\, dr. 
\end{equation}
For the specific case of Tolman~IV we can write 
\begin{eqnarray}
    F_{eq}
    &>&  \frac{1}{4 }\lint^{\sqrt{\frac{R^2 - A^2}{3}}}_0 \frac{R^2 - A^2 - 3r^2}{R^2(A^2 + 2r^2)} r dr.
\end{eqnarray}
Now make the substitutions $y= r^2/R_s^2$ and $a=A^2/R^2$. We find
\begin{equation}
F_{eq} > \pi r_s^2 \int^{1}_0p\; dy =
{1\over 8} \int^{1}_0 \frac{\left(1-a\right)^2  \left(1-y\right)}{2 \left(1-a\right) y+3 a} \; dy. 
\end{equation}
This integral yields
\begin{equation}
F_{eq} > \left. \left\{{(a+2)\ln[(2y-3)a-2y]\over32}  -{(1-a)y\over16} \right\} \right|_0^1. 
\end{equation}
Thence
\begin{equation}
F_{eq} > {(a+2)[\ln(2+a)-\ln(3a)\over32}  -{(1-a)\over16}. 
\label{eq:F_eqInequal}
\end{equation}
Under the limit $a\rightarrow0$ we find that the term $-\log(3a) \rightarrow +\infty$. So the inequality  (\ref{eq:F_eqInequal}) diverges to infinity, demonstrating that the equatorial force in the Tolman-IV space-time can be made to violate the weak maximum force conjecture.

Thus, as in the case of the interior Schwarzschild solution, we have shown that the radial force is bounded (and in this case obeys both the weak and strong maximum force conjectures). However, the equatorial force can be made to diverge to infinity and act as a counter example to both weak and strong conjectures.

\subsubsection{DEC}

To see if the DEC is satisfied over the range of integration for the equatorial force, we inquire as to whether or not
\begin{equation}
    \frac{p}{\rho} =
\frac{(A^2 + 2r^2)(R^2 - A^2 - 3r^2)}{3A^4 + 7A^2r^2 + 6r^4 + (3A^2 + 2r^2)R^2} \leq1\;?
\end{equation}
It is straightforward to check that this inequality will always hold in the physical region.
Using the definitions $a=A^2/R^2$ and $z=r^2/R^2$, so that $a\in(0,1)$, 
and $z\in(0,{1-a\over3})$, we can write this as
\begin{equation}
{p\over\rho}-1 = 
-{2(2a^2+6az+a+6z^2)\over(7a+2)z +3a(a+1) +6z^2} 
<0,
\end{equation}
which is manifestly negative. 
So adding the DEC does not affect or change our conclusions.
Indeed, we have already seen that the equatorial force diverges in the limit of $a\to0$ implying $A\rightarrow 0$. Applying this limit to the ratio $p/\rho$ gives:
\begin{equation}
    \lim_{A \rightarrow 0 } \frac{p}{\rho} = 1 - \frac{6r^2}{3r^2+R^2}= 1 - {6y\over1+3y} \leq 1.
\end{equation}
Again, this is always true within any physical region,
so we verify that adding the DEC does not change our conclusions.

\subsubsection{Summary}

For the Tolman IV solution, while the radial force is bounded (and  obeys both the weak and strong maximum force conjectures), the equatorial force can be made to diverge to infinity in certain parts of parameter space ($A\to0$) and acts as a counter-example to both weak and strong maximum force conjectures. For the Tolman IV solution, adding the DEC does not save the situation, the violation of both weak and strong  maximum force conjectures is robust.

\subsection{Buchdahl--Land spacetime: $\rho = \rho_s + p$}
The Buchdahl--Land spacetime is a special case of the Tolman IV spacetime, corresponding to the limit  $A\rightarrow0$ (equivalently $a\rightarrow0$).  It is sufficiently simple that it is worth some discussion in its own right. The Tolman IV metric (with a re-scaled time coordinate $t\rightarrow At$) can be written:
\begin{equation}
    ds^2 = -(A^2 + r^2)dt^2 + \frac{1 + \frac{2r^2}{A^2}}{\left(1 - \frac{r^2}{R^2}\right)\left(1 + \frac{r^2}{A^2}\right)}dr^2 + r^2d\Omega^2.
\end{equation}
Under the limit $A\rightarrow0$, this becomes:
\begin{equation}
    ds^2 = -r^2dt^2 + \frac{2R^2}{R^2-r^2}dr^2 + r^2d\Omega^2.
    \label{eq:A=0 Limit TIV}
\end{equation}
Then the orthonormal stress-energy components are:
\begin{eqnarray}
T_{\hat{t}\hat{t}} &=& \rho = \frac{1}{16\pi}\left(\frac{1}{r^2}+\frac{3}{R^2}\right); \nonumber\\
T_{\hat{r}\hat{r}} &=& T_{\hat{\theta}\hat{\theta}} = T_{\hat{\phi}\hat{\phi}} = p = \frac{1}{16\pi}\left(\frac{1}{r^2}-\frac{3}{R^2}\right).
\end{eqnarray}
The surface is located at
\begin{equation}
    R_s = \frac{R}{\sqrt{3}};   \qquad \hbox{with} \qquad \rho_s = {3\over8\pi R^2} = {1\over8\pi R_s^2}.
\end{equation}
At the centre the pressure and density both diverge --- more on this point later.

We recast the metric as
\begin{equation}
    ds^2 = -r^2dt^2 + {2 \over1-{1\over3}{r^2\over R_s^2}} dr^2 + r^2d\Omega^2.
    \label{eq:BuchLand}
\end{equation}
This is simply a relabelling of equation (\ref{eq:A=0 Limit TIV}).
The orthonormal stress-energy tensor is now relabelled as:
\begin{eqnarray}
T_{\hat{t}\hat{t}} &=& \rho = \frac{1}{16\pi}\left(\frac{1}{r^2} +{1\over R_s^2} \right);\nonumber\\
T_{\hat{r}\hat{r}} &=&  T_{\hat{\theta}\hat{\theta}} = T_{\hat{\phi}\hat{\phi}} = p = \frac{1}{16\pi}\left(\frac{1}{r^2} -{1\over R_s^2}\right).
\end{eqnarray}
Note that
\begin{equation}
p = \rho-\rho_s; \qquad \hbox{that is} \qquad \rho=\rho_s + p.
\end{equation}
That is, the Buchdahl--Land spacetime represents a ``stiff fluid''. 
This perfect fluid solution has a naked singularity at $r=0$ and a well behaved surface at finite radius.
The singularity at $r=0$ is not really a problem as one can always excise a small core region near $r=0$ to regularize the model.

\subsubsection{Radial force}
Due to the simplicity of the pressure, the radial force can be easily calculated as:
\begin{equation}
    F_r =  \frac{1}{4}\left(1 -{r^2\over R_s^2}\right).
\end{equation}
The radial force is trivially bounded with a maximum of $\frac{1}{4}$ at the centre of the star. 
This  obeys the strong (and so also the weak) maximum force conjecture.

\subsubsection{Equatorial force}
The equatorial force is:
\begin{eqnarray}F_{eq} &=& 2\pi\lint^{{R_s}}_{\!\!\!\!\!0}
\sqrt{ 2 \over1-{1\over3}{r^2\over R_s^2}}\;\;
\frac{1}{16\pi}\left(\frac{1}{r^2} -{1\over R_s^2}\right)\; r dr.
\end{eqnarray}
This is now simple enough to handle analytically. Using the dimensionless variable $y= r^2/R_s^2$, with range $y\in(0,1)$, we see:
\begin{eqnarray}F_{eq} &=& {1\over16} \lint^{1}_{\!\!\!\!\!0}
\sqrt{ 2 \over1-{1\over3}{y}}\;\;
\left(1-y\right)\; {dy\over y}.
\end{eqnarray}
This is manifestly dimensionless, and manifestly diverges to $+\infty$. 
If we excise a small region  $r<r_\mathrm{core}$, (corresponding to $y < y_\mathrm{core}$) to regularize the model, replacing $r<r_\mathrm{core}$
with some well-behaved fluid ball, then we have the explicit logarithmic divergence
\begin{equation}
F_{eq} = -{1\over16} \ln y_\mathrm{core} + \O(1).
\end{equation}
This  violates the weak (and so also the strong) maximum force conjecture.

\subsubsection{DEC}
The DEC for this space-time is given by:
\begin{eqnarray}
\frac{p}{\rho} = {\rho-\rho_s\over \rho} = 1 - {\rho_s\over\rho} \leq 1.
\end{eqnarray}
which is always true for positive values of $r$, $\rho_s$. 

\subsubsection{Summary}

The Buchdahl--Land spacetime is another weak maximum force conjecture counter-example, one which again obeys the classical energy conditions.

\subsection{Scaling solution}
The scaling solution is
\begin{equation}
    ds^2 = -r^{\frac{4w}{1 + w}} dt^2 + \left(\frac{w^2 + 6w + 1}{(1+w)^2}\right) dr^2 + r^2d\Omega^2.
\end{equation}
This produces the following stress energy tensor:
\begin{eqnarray}
T_{\hat{t}\hat{t}} &=& \rho = \frac{w}{2\pi(w^2 + 6w + 1)r^2};\nonumber\\
T_{\hat{r}\hat{r}} &=& T_{\hat{\theta}\hat{\theta}} = T_{\hat{\phi}\hat{\phi}} = p = \frac{w^2}{2\pi(w^2 + 6w + 1)r^2}.
\end{eqnarray}
This perfect fluid solution has a naked singularity at $r=0$ and does not have a finite surface --- it  requires $r\rightarrow\infty$ for the pressure to vanish. Nevertheless, apart from a small region near $r=0$ and small fringe region near the surface $r=R_s$, this is a good approximation to the bulk geometry of a star that is on the verge of collapse~\cite{collapse,Yunes}. 
To regularize the model excise two small regions, a core region at $r\in(0,r_\mathrm{core})$, and an outer shell at $r\in(r_\mathrm{fringe},R_s)$, replacing them by segments of well-behaved fluid spheres.
Note that for $r\in(r_\mathrm{core},r_\mathrm{fringe})$ we have $p/\rho=w$, (and since $\rho>0$ we must have $w>0$), so the DEC implies $w\in(0,1]$. 

\subsubsection{Radial force}
Using equation (\ref{eq:Radial Force}), we find that the radial force is very simply given by:
\begin{equation}
    F_r = \frac{2w^2}{w^2 + 6w + 1}.
\end{equation}
This is independent of $r$ and attains a maximum value of $\frac{1}{4}$ when $w=1$, giving a bounded force obeying the strong maximum force conjecture.

\subsubsection{Equatorial force}
Now, using equation (\ref{eq:Equatorial Force2}), the equatorial force can be calculated as:
\begin{equation}
    F_{eq} =\lint^{r_\mathrm{fringe}}_{\!\!\!\!\!r_\mathrm{core}}\sqrt{\frac{w^2 + 6w + 1}{(1+w)^2}}\left(\frac{w^2}{(w^2 + 6w + 1)r}\right) dr + \O(1).
\end{equation}
That is
\begin{equation}
    F_{eq} =\frac{w^2}{(1+w)\sqrt{w^2 + 6w + 1}}\;
    \ln(r_\mathrm{fringe}/r_\mathrm{core}) +\O(1),
\end{equation}
which trivially diverges logarithmically as either $r_\mathrm{core}\to0$ or $r_\mathrm{fringe}\to\infty$,
providing a  counter-example to weak maximum force conjecture.

\subsubsection{Summary}

Again we have an explicit model where the radial force $F_r$ is well-behaved, but the equatorial force $F_{eq}$ provides an explicit counter-example to weak maximum force conjecture.
This counter-example is compatible with the DEC.

\subsection{TOV equation}
Let us now see how far we can push this sort of argument using only the TOV equation for the pressure profile in perfect fluid spheres --- we will (as far as possible) try to avoid making specific assumptions on the metric components and stress-energy. The TOV equation is
\begin{equation}
{dp(r)\over dr} = - {\{\rho(r)+p(r)\}\{m(r)+4\pi p(r) r^3\}\over r^2\{1-2m(r)/r\}}.
\end{equation}

\subsubsection{Radial force}
From the definition of radial force $F_r = 4\pi p r^2$, we see that at the maximum  of $F_r$ we must have 
\begin{equation}
\left.(2 p r + r^2 p')\right|_{r_\mathrm{max}}=0.
\end{equation}
Thence, at the maximum
\begin{equation}
(F_r)_\mathrm{max} = (4\pi p r^2)_\mathrm{max} = -2\pi \left.(r^3 p')\right|_{r_\mathrm{max}}.
\end{equation}
In particular, now using the TOV at the location $r_\mathrm{max}$ of the maximum  of $F_r$:
\begin{equation}
(F_r)_\mathrm{max} = 
2\pi \left.\left[ {(\rho+p) r(m +4\pi p r^3)\over (1-2m/r)}\right]\right|_{r_\mathrm{max}}.
\end{equation}
Let us define the two parameters 
\begin{equation}
\chi=\left[2m(r)\over r\right]_{r_\mathrm{max}}={2m(r_\mathrm{max})\over r_\mathrm{max}},
\quad\hbox{and}\quad
w = \left[p(r)\over\rho(r)\right]_{r_\mathrm{max}}= {p(r_{max})\over\rho(r_{max})}.
\end{equation}
Then
\begin{equation}
(F_r)_\mathrm{max}  = 
{1\over2} \left.\left({4\pi p (1+{1\over w}) r^2(\chi/2 +4\pi p r^2)\over 1-\chi}\right)\right|_{r_\mathrm{max}}.
\end{equation}
Simplifying, we see:
\begin{equation}
(F_r)_\mathrm{max}  = 
{1\over2} {(F_r)_\mathrm{max}\; [1+1/w]\; [(F_r)_\mathrm{max}+\chi/2)\over 1-\chi}
\end{equation}
Discarding the unphysical solution $(F_r)_\mathrm{max}  = 0$, we find
\begin{equation}
(F_r)_\mathrm{max}  = {4w -\chi-5w\chi\over2(1+w)} = {2w\over1+w} - \chi \;{1+5w\over1+w}.
\end{equation}
The physical region corresponds to $0\leq \chi <1$, while $w>0$.  Furthermore we have  $(F_r)_\mathrm{max} >0$, whence $4w -\chi-5w\chi>0$, implying $\chi < 4w/(1+5w) < 4/5$. That is, at the location $r_\mathrm{max}$  of the maximum of $F_r$ we have
\begin{equation}
\left[2m(r)\over r\right]_{r_\mathrm{max}}  = {2m(r_\mathrm{max})\over r_\mathrm{max}} < {4\over5}.
\end{equation}
This is not the Buchdahl--Bondi bound, it is instead a bound on the compactness of the fluid sphere at the internal location $r_\mathrm{max}$ where $F_r$ is maximized.

Observe that $(F_r)_\mathrm{max} $ is maximized when $\chi=0$ and $w=\infty$, when $(F_r)_\mathrm{max}\to 2$. This violates the strong conjecture maximum force but not the weak maximum force conjecture. 
If we impose the DEC then $w\leq 1$, and $(F_r)_\mathrm{max} $ is maximized when $\chi=0$ and $w=1$, when $(F_r)_\mathrm{max}\to 1$. This still violates the strong maximum force conjecture but not the weak maximum force conjecture. 
Consequently the weak conjecture for $F_r$ generically holds for any prefect fluid sphere satisfying the TOV.

\subsubsection{Equatorial force}
As we have by now come to expect, dealing with the equatorial force will be considerably trickier. 
In view of the non-negativity of the Misner--Sharp quasi-local mass we have:
\begin{equation}
F_{eq} = 2\pi \int_0^{R_s} \sqrt{g_{rr}} \; p \; r dr
= 2\pi \int_0^{R_s} {1\over\sqrt{1-2m(r)/r}} \; p \; r dr > 2\pi\int_0^{R_s} p \, r \; dr .
\end{equation}
To make the integral $\int_0^{R_s} p \, r \; dr $ converge it is sufficient to demand $p(r) = o(1/r^2)$.
However, for stars on the verge of gravitational collapse it is known that $p(r) \sim K/r^2$, see for instance~\cite{collapse, Yunes}. More specifically, there is some core region $r\in(0,r_\mathrm{core})$ designed to keep the central pressure finite but arbitrarily large, 
a large scaling region $r\in (r_\mathrm{core},r_\mathrm{fringe})$ where $p \sim K/r^2$, and an 
outer fringe $r\in (r_\mathrm{fringe},R_s)$ where one has $p(r)\to p(R_s)=0$. 
Then we have the identity
\begin{equation}
\int_0^{R_s} p \, r \; dr  = \int_0^{r_\mathrm{core}} p \, r \; dr  + \int_{r_\mathrm{core}}^{r_\mathrm{fringe}} p \, r \; dr + \int_{r_\mathrm{fringe}}^{R_s} p \, r \; dr.
\end{equation}
But under the assumed conditions this implies
\begin{equation}
\int_0^{R_s} p \, r \; dr  = \O(1) + 
\left[\int_{r_\mathrm{core}}^{r_\mathrm{fringe}} {K\over r} \; dr + \O(1) \right] 
+\O(1).
\end{equation}
Thence
\begin{equation}
\int_0^{R_s} p \, r \; dr  = K \ln\left({r_\mathrm{fringe}}/r_\mathrm{core} \right)
+\O(1).
\end{equation}
Finally
\begin{equation}
F_{eq} > 2\pi K \ln\left({r_\mathrm{fringe}}/r_\mathrm{core} \right)
+\O(1).
\end{equation}
This can be made arbitrarily large for a star on the verge of gravitational collapse, so the weak and strong maximum force conjectures are both violated. 

Note that technical aspects of the argument are very similar to what we saw for the exact scaling solution to the Einstein equations, but the physical context is now much more general. 

\subsubsection{Summary}

We see that the weak maximum force conjecture generically holds for the radial force $F_r$ when considering perfect fluid spheres satisfying the TOV. In contrast we see that the weak maximum force conjecture fails for the equatorial force $F_{eq}$ when considering perfect fluid spheres satisfying the TOV that are close to gravitational collapse. 

\section{Discussion}
With the notion a natural unit of force $F_* = F_\mathrm{Planck} = c^4/G_N$ in hand, one can similarly define a natural unit of power~\cite{Dyson, Barrow:2017, footnote5, Cardoso:2018, Bruneton:2013}
\begin{equation}
P_*= P_\mathrm{Planck} = {c^5\over G_N} = 1 \hbox{ Dyson} \approx 3.6\times 10^{52} \hbox{ W},
\end{equation}
a natural unit of mass-loss-rate
\begin{equation}
 (\dot m)_* = (\dot m) _\mathrm{Planck}= {c^3\over G_N} \approx 4.0\times 10^{35} \hbox{ kg/s},
\end{equation}
and even a natural unit of mass-per-unit-length
\begin{equation}
 (m')_* = (m') _\mathrm{Planck}= {c^2\over G_N} \approx 1.36\times 10^{27} \hbox{ kg/m}. 
\end{equation}
Despite being Planck units, all these concepts are purely classical (the various factors of $\hbar$ cancel, at least in (3+1) dimensions). 

Indeed, consider the classical Stoney units which pre-date Planck units by some 20 years~\cite{Stoney1, Stoney2, Stoney3}, and use $G_N$, $c$,
and Coulomb's constant $e^2\over4\pi\epsilon_0$, instead of $G_N$, $c$, and Planck's constant $\hbar$. 
Then we have $F_* = F_\mathrm{Planck} =F_\mathrm{Stoney}$. Similarly  we have $P_*= P_\mathrm{Planck}= P_\mathrm{Stoney}$,  $(\dot m)_* = (\dot m)_\mathrm{Planck}= (\dot m)_\mathrm{Stoney}$, 
and $(m')_* = (m') _\mathrm{Planck} = (m')_\mathrm{Stoney}$. 
Based ultimately on dimensional analysis, any one of these quantities might be used to advocate for a maximality conjecture: maximum luminosity~\cite{Dyson,  Barrow:2017, footnote5, Cardoso:2018, Bruneton:2013}, maximum mass-loss-rate, or maximum mass-per-unit-length. 
The specific counter-examples to the maximum force conjecture that we have discussed above suggest that it might also be worth looking for specific counter-examples to these other conjectures~\cite{Cardoso:2018}. 

\section{Conclusions}

Through the analysis of radial and equatorial forces within perfect fluid spheres in general relativity, we have  produced a number of counter-examples to both the strong and weak forms of the maximum force conjecture. These counter-examples highlight significant issues with the current phrasing and understanding of this conjecture, as merely specifying that forces are bounded within the framework of general relativity is manifestly a falsehood. 
As such, should one wish some version of the maximum force conjecture to be considered viable as a potential physical principle, it must be very clearly specified as to what types of forces they pertain to.

\section*{Acknowledgments}
MV was supported by the Marsden Fund, via a grant administered by the Royal Society of New Zealand.


\end{document}